\documentclass[10pt]{article}

\usepackage{amsmath}
\usepackage{amssymb}

\usepackage{graphicx}
\usepackage{caption}
\usepackage{subfigure}
\usepackage{cite}

\usepackage{color} 

\usepackage[labelfont=bf,labelsep=period,justification=raggedright]{caption}

\makeatletter
\renewcommand{\@biblabel}[1]{\quad#1.}
\makeatother
\date{}

\begin{document}

\begin{flushleft}
{\Large
\textbf{Brain Network Adaptability Across Task States}
}

Elizabeth N. Davison$^{1}$,
Kimberly J. Schlesinger$^{1}$,
Danielle S. Bassett$^{2, 3}$,
Mary-Ellen Lynall$^{4}$, 
Michael B. Miller$^{5}$,  
Scott T. Grafton$^{5}$,  
Jean M. Carlson$^{1,\ast}$
\\
\bf{1} Department of Physics, University of California, Santa Barbara, CA 93106 USA
\\
\bf{2} Department of Bioengineering, University of Pennsylvania, Philadelphia, PA 19104 USA
\\
\bf{3} Department of Electrical and Systems Engineering, University of Pennsylvania, Philadelphia, PA 19104 USA
\\
\bf{4} Department of Medical Sciences, University of Oxford, Oxford OX1 3LB, UK
\\
\bf{5} Department of Psychological and Brain Sciences, University of California, Santa Barbara, CA 93106
\\
$\ast$ E-mail: Corresponding carlson@physics.ucsb.edu
\end{flushleft}

\section*{Abstract}
Activity in the human brain moves between diverse functional states to meet the demands of our dynamic environment, but fundamental principles guiding these transitions remain poorly understood. Here, we capitalize on recent advances in network science to analyze patterns of functional interactions between brain regions. We use dynamic network representations to probe the landscape of brain reconfigurations that accompany task performance both within and between four cognitive states: a task-free resting state, an attention-demanding state, and two memory-demanding states. Using the formalism of hypergraphs, we identify the presence of groups of functional interactions that fluctuate coherently in strength over time both within (task-specific) and across (task-general) brain states. In contrast to prior emphases on the complexity of many dyadic (region-to-region) relationships, these results demonstrate that brain adaptability can be described by common processes that drive the dynamic integration of cognitive systems. Moreover, our results establish the hypergraph as an effective measure for understanding functional brain dynamics, which may also prove useful in examining cross-task, cross-age, and cross-cohort functional change.

\section*{Author Summary}
The human brain is a complex system in which the interactions of many billions of neurons give rise to a fascinating range of human behaviors. In response to its changing environment -- for example, in situations involving rest, memory, focused attention, or learning -- the brain dynamically switches between distinct patterns of functional activation. Despite the wealth of neuroimaging data available, the immense complexity of the brain makes the identification of fundamental principles guiding this task-based organization of neural activity a distinct challenge. Here, we apply new techniques from dynamic network theory to describe interactions between brain regions as an evolving network; this provides a framework for understanding these time-dependent interactions in terms of organizing characteristics of the whole network. We examine patterns of correlated neural activity during four states: a resting state, an attention-demanding state, and two memory-demanding states. Using network science techniques, we identify groups of brain network interactions that change cohesively together over time, both within individual tasks and across all tasks. By developing tools to analyze the size and spatial distributions of these groups, we quantify significant differences between properties of brain network dynamics in different tasks. This work demonstrates that these techniques quantitatively characterize evolving task-based organization of neural activity and provide a promising method for investigating how brain network properties of individuals correspond to performance on cognitive tasks.

\section*{Introduction}
An essential characteristic of the human brain is the ability to transition between functional states in synchrony with changing demand. A central focus in neuroscience involves quantifying this adaptability by delineating changes in regional blood-oxygen-level-dependent (BOLD) signal associated with different cognitive tasks, or between task states and task-free (resting \cite{Raichle2001, Damoiseux2006}) states \cite{Fox2005,Hampson2010}. However, this approach, which examines the magnitude of brain activity alone, is unable to completely describe the correlation structure linking spatially segregated neural circuits.

Recent advances in network science provide tools to represent and characterize the functional interactions between brain regions forming cognitive systems. In this formalism, brain regions are represented as network nodes and functional connections (estimated by statistical similarities between BOLD signals \cite{friston_functional_2011}) are represented as network edges \cite{Bassett2006a,Bullmore2009}. These approaches enable the statistically principled examination of large-scale neural circuits underlying cognitive processes. Moreover, they have enabled quantitative comparisons \emph{between} circuits \cite{Mennes2010,Cole2013}.

A critical open question remains: Do functional correlations evolve independently within and across brain states? Or do they evolve in sets, each set controlled by a common regulatory driver? To answer this question, we estimate brain functional networks in one minute time intervals as 86 participants engaged in four task states: a task-free resting state, an attention-demanding state, and two memory-demanding states. We treat the evolving patterns of functional connectivity as temporal networks \cite{Bassett2011b,Bassett2012b,Bassett2013a,mantzaris_dynamic_2013,Doron2012,Sieben2013a} and estimate the pairwise correlation between the strengths of functional interactions over time. We demonstrate the existence of groups of functional interactions (\emph{hyperedges} \cite{bassett_cross-linked_2013}) which display similar changes in strength within and across task states. We characterize hyperedges within and across task states according to their structure, anatomy, and task specificity. These analyses provide a unique and novel window into the adaptability of the brain as it transitions between states and offer quantitative statistics for the comparison of such adaptability across subject cohorts.

\section*{Methods}
\subsection*{Imaging}
MRI data was acquired at the UCSB Brain Imaging Center from 116 healthy adult participants using a phased array 3T Siemens TIM Trio with a 12 channel head coil. Functional MRI data was taken while each participant engaged in the following four tasks: rest (task-free), two separate functional runs of the same attention-demanding task, a memory task with lexical stimuli, and a memory task with face stimuli. In both memory tasks, 180 previously studied stimuli and 180 novel stimuli were presented to the subjects, who were asked to determine whether each stimulus was ``old" or ``new" -- i.e., whether it had been previously studied. This analysis combines two separate functional runs of the same attention task, which required the subjects to detect the presence or absence of a target stimulus \cite{Hermundstad2013a}. During the resting state, participants were asked to lie still and look at a blank screen. The sampling period (TR) was 2s for the rest and attention tasks and 2.5s for both memory tasks. In addition to functional data, a three dimensional high-resolution T1-weighted structural image of the whole brain was obtained for each participant. For additional experimental details, see \cite{Hermundstad2013a},\cite{aminoff_individual_2012}, and supplemental information therein.

\subsection*{Image Analysis}

\subsubsection*{Structural MRI acquisition and pre-processing}
Structural scans were intensity-corrected, skull-stripped, normalized, segmented and parcellated using Freesurfer v.5.0.0 cortical reconstruction all with default settings, accessed via the Connectome Mapping Toolkit v.1.2.0 \cite{daducci2012connectome}. The starting atlas was the updated Lausanne2008 multi-scale atlas \cite{hagmann2008mapping}. For each subject, parcellations containing 83, 129, 234, 463 and 1015 regions were generated, covering cortical grey-matter regions, the thalamus, caudate, putamen, pallidum, accumbens area, hippocampus, amygdala and brainstem. The highest-resolution parcellation of 1015 regions was not investigated further, since a large number of regions contained very few or no voxels when the atlas was downsampled into fMRI space.

\subsubsection*{Functional MRI pre-processing and time series analysis}
Preprocessing was performed using FSL v5.0 \cite{smith2004advances,woolrich2009bayesian,jenkinson2012fsl}, AFNI v. 2011 12 21 1014 http://afni. nimh.nih.gov \cite{cox1996anfi} and Matlab (2013, The Mathworks, Natick, MA). Functional MRI scans were preprocessed as follows. FSL programs MCFLIRT \cite{jenkinson2002improved} and fsl motion outliers were used to correct for head motion and derive a volume-by-volume measure of head motion: framewise displacement. Framewise displacement (FD) is calculated as the sum (in mm) of rotational and translational displacements from volume N to N+1 \cite{power2012spurious}. Next, we performed slice timing correction (AFNI 3dTshift), auto-masked to obtain a brain-only fMRI image (AFNI 3dAutomask), and smoothed the time series at each voxel (AFNI 3dDespike with default parameter settings). Despiking has been shown to reduce the motion-related distance dependent bias in correlation estimates \cite{jo2013effective}. Each voxel's time series was then detrended with respect to framewise displacement using AFNI 3dDetrend. This uses linear regression to remove variability related to the nuisance regressor, framewise displacement, at each voxel. Runs were only included in the analysis if mean framewise displacement for the run was less than 0.25mm per frame; this led to 73 fMRI runs (of 763 total runs) being excluded from this analysis. Registration proceeded as follows: a participant's time-averaged fMRI image was aligned to their structural T1 scan using FSL FLIRT boundary-based registration \cite{greve2009accurate,jenkinson2002improved}, and the inverse of this transformation was applied to all subjects’ parcellation scales (generated in structural space). Parcellations were downsampled into EPI (AFNI 3dfractionize, voxel centroid voting, requiring 60\% overlap), and the mean signal across all the voxels within a given brain region was calculated to produce a single representative time series. The data was not spatially smoothed at any stage.

\subsubsection*{Creation of a hybrid atlas}
We sought to create an atlas with low inter-individual and cross-brain variability in the amount of fMRI data acquired per region. Many existing atlases use parcellations that have roughly equal region sizes as measured on structural MRI scans \cite{zalesky2010whole}. However, downsampling the atlas from structural MRI voxels to fMRI voxels, along with inhomogeneous fMRI signal-loss, mean that this does not produce equally sized regions in functional MRI space. To mitigate this, we generated a `hybrid', atlas by choosing those regions from various scales of the Lausanne2008 atlas that minimized cross-brain and intra-subject variability in region size. Starting with the scale 234 atlas, an iterative process was used to decrease intra- and intersubject variability in region size. Where a region had very few voxels (mean size $<$ 25th percentile), or high variability in size across subjects (coefficient of variation $>$ 30\%), it was tentatively exchanged for a region from the next highest resolution atlas, effectively combining the initial region with other higher-resolution regions subsumed under the same anatomical heading. If this combination of regions decreased the inter-subject or within-subject variability in region size, the combined region was retained. If not, the initial poor quality region was rejected from the ``hybrid atlas''. This was repeated until no further combinations of regions could decrease intra- and inter-subject variability while retaining neuroanatomically sensible groupings. Regions were excluded from the analysis altogether if there were fMRI runs in which no data was acquired in that region (frontal pole, entorhinal cortex and temporal pole), or if the inter-subject coefficient of variation was greater than 30\% (this applied to 7 of the 8 inferior temporal regions; 1 of the 8 middle temporal regions; 2 of 8 fusiform regions; 1 of the 6 caudal middle frontal regions, and 1 of the 14 precentral regions). Table \ref{Table 1} lists the 194 regions identified by this hybrid atlas. This approach considerably reduced intra-subject variability in region size as well as reducing the inter-subject variability at problematic outlier regions, while minimizing the amount of data that had to be excluded from analysis. 

\begin{table}[!ht]
\vspace{5 mm}
\begin{center}
\begin{tabular}{ | l | l | l | p |}
           \hline
           \textbf{Region Name} & \textbf{L} & \textbf{R} \\ \hline
     lateralorbitofrontal & 2 & 2\\ \hline
     parsorbitalis & 1 & 1 \\ \hline
     medialorbitofrontal & 1 & 1\\ \hline
     parstriangularis & 1 & 1\\ \hline
     parsopercularis & 2  & 2  \\ \hline
     rostralmiddlefrontal & 5  & 6  \\ \hline
     superiorfrontal & 9  & 8  \\ \hline
     caudalmiddlefrontal & 3 & 2 \\ \hline
     precentral & 7 & 6 \\ \hline
     paracentral & 1 & 1 \\ \hline
     rostralanteriorcingulate & 1 & 1 \\ \hline
     caudalanteriorcingulate & 0 & 1 \\ \hline
     posteriorcingulate & 2 & 2 \\ \hline
     isthmuscingulate & 1 & 1 \\ \hline
     postcentral & 7 & 5 \\ \hline
     supramarginal & 5 & 4 \\ \hline
     superiorparietal & 7 & 7 \\ \hline
     inferiorparietal & 5 & 6 \\ \hline
     precuneus & 5 & 5 \\ \hline
      \end{tabular}
      \hspace{1cm}
      \begin{tabular}{ | l | l | l | p |}
 	\hline
 	\textbf{Region Name} & \textbf{L} & \textbf{R} \\ \hline
 	    cuneus & 1 & 1 \\ \hline
 	pericalcarine & 1 & 1 \\ \hline
 	lateraloccipital & 5 & 5 \\ \hline
 	lingual & 2 & 3 \\ \hline
 	fusiform & 3 & 3 \\ \hline
 	parahippocampal & 1 & 1 \\ \hline
 	inferiortemporal & 1 & 0 \\ \hline
 	middletemporal & 3 & 4 \\ \hline
 	bankssts & 1 & 1 \\ \hline
 	superiortemporal & 5 & 5 \\ \hline
 	transversetemporal & 1 & 1 \\ \hline
 	insula & 2 & 2 \\ \hline
 	thalamusproper & 1 & 1 \\ \hline
 	caudate & 1 & 1 \\ \hline
 	putamen & 1 & 1 \\ \hline
 	pallidum & 1 & 1 \\ \hline
 	accumbensarea & 1 & 1 \\ \hline
 	hippocampus & 1 & 1 \\ \hline
 	amygdala & 1 & 1 \\ \hline
        \end{tabular}
        \end{center}
\caption{
{\bf Brain Regions} The 194 regions used in the hyperedge analysis, listed in order and including the number of regions in left and right hemispheres.}
 \label{Table 1}
 \end{table}
 
\subsection*{Functional Connectivity}
Specific frequencies of oscillations in the BOLD signal have been associated with different cognitive functions. We focus our investigation on low frequency (0.06-0.125 Hz) oscillations in the BOLD signal that have proven useful for examining resting \cite{Lynall2010,Bassett2012a} and task-based functional connectivity \cite{Bassett2011b}. The task-related oscillations are posited to be specific to this frequency range, possibly due to a bandpass-filter-like effect from the hemodynamic response function \cite{sun_measuring_2004}. We apply a Butterworth bandpass filter to isolate frequencies in the (0.06-0.125 Hz) range \cite{cadzow_discrete-time_????}.

To construct a functional brain network, we use the 194 region hybrid atlas, where each region contains a roughly equal number of voxels. These 194 regions represent the network nodes. The $x$, $y$, and $z$ positions of each node are given by the centroid of the voxels which comprise the node. Edge weights in the functional brain network are computed by taking Pearson's correlations between the filtered time series within a defined time period for each pair of nodes \cite{fornito_network_2010}.
 
\subsection*{Time Windows for Temporal Network Construction} 
Dynamic networks are constructed by taking the filtered time series in temporal windows of 60 seconds and computing a $N \times N$ adjacency matrix of nodal correlations for each time window, where $N=194$ is the number of nodes. Each of these $N\times N$ adjacency matrices represents the functional network over the 60 seconds in question. From this set of networks, we extract the edge weight time series by considering the correlation strength in each sequential network. We let $E = N(N-1)/2=18721$ be the total number of edges between the 194 nodes and construct an $E \times E$ adjacency matrix \textbf{X}, where $X_{ab}$ gives the Pearson correlation coefficient between the time series of edge weight for edges $a$ and $b$. The entries of the $E\times E$ adjacency matrix represent pairs of edges with correlated weight time series \cite{bassett_cross-linked_2013}. We consider a range of temporal window lengths from 40 to 120 seconds and find that our results for hyperedge size and spatial distributions are robust to changes in window length in this range.

\subsection*{Hyperedge Construction}
\begin{figure}[!ht]
\centering
\includegraphics[width = 0.75\textwidth]{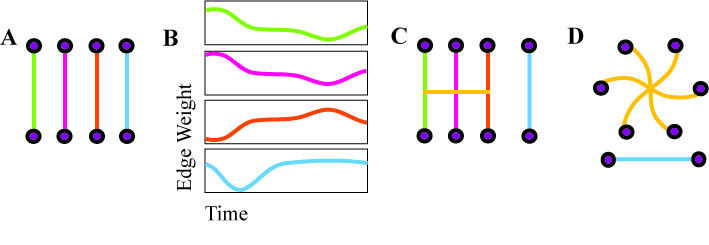}
\caption{\textbf{Hyperedge Construction:} A schematic illustration of the method used to identify hyperedges. We begin with a set of node-node edges (A) and their time series (B), of which three [green, pink and orange traces, (B)] exhibit strong pairwise temporal correlations. These edges are cross-linked (C) by temporal covariance in edge weight time series, and thereby form a hyperedge (D) of size three on six nodes. The final [blue] edge forms a singleton, an edge which is not significantly correlated with any other edges.}
\label{Figure 1}
\end{figure}

The cross-linked network structure, which contains information about groups of edges with similar time series (hyperedges), is extracted from the edge-edge correlation matrix \textbf{X} \cite{bassett_cross-linked_2013}. Figure \ref{Figure 1} provides a schematic illustration of the process of determining the cross-linked structure of a network. To exclude entries of \textbf{X} that are not statistically significant, we threshold \textbf{X} by evaluating the $p$-values for the Pearson coefficient $R$ for each edge-edge correlation using a false discovery rate correction for false positives due to multiple comparisons \cite{FDR}. If the $p$-value for an entry $X_{ij}$ satisfies the false discovery rate correction threshold, we set $\xi_{ij} = R(i,j)$ for our thresholded matrix $\xi$. We set the thresholded entry of all other elements $X_{ij}$ to zero. We binarize this thresholded matrix and obtain $\xi_{ij}^\prime$, where
\begin{align}
\xi_{ij}^\prime = \begin{cases}
1, & \text{if} \;\;\xi_{ij} >0 ;\\ 0, & \text{if} \;\;\xi_{ij} = 0.
\end{cases}
\end{align}

\begin{figure}[!ht]
 \centering
 \includegraphics[width = 0.5\textwidth]{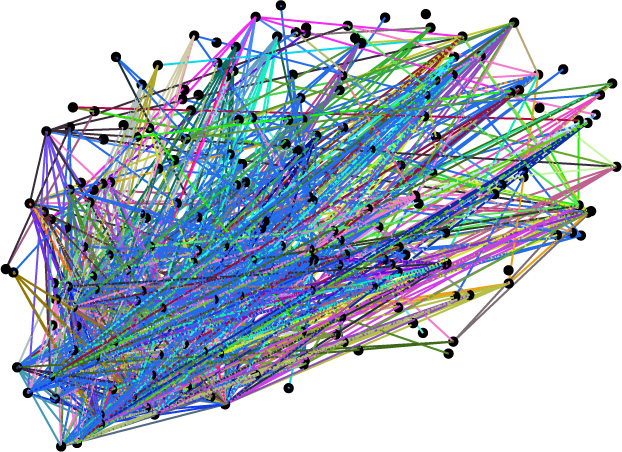}
 \caption{\textbf{Individual Hypergraph:} Hypergraph for Subject 68, chosen for visualization due to the relatively small maximum hyperedge size of 256. In this representation, different colors represent separate hyperedges, and only edges of size four or greater are shown. A hypergraph with maximum hyperedge size of thousands of edges is difficult to visualize because it obscures any structure attributed to the smaller hyperedges.}
 \label{HI}
 \end{figure}

Each connected component in $\xi$ represents a hyperedge, a set of edges that have significantly correlated temporal profiles. The groups of nodes in Figure \ref{Figure 1} (D) are examples of such connected components. The set of all hyperedges defined in $\xi$ produces an individual hypergraph, illustrated for one subject in Figure \ref{HI}. These hyperedges contain information about edge dynamics without restricting the analysis to edges with strong weights.

\subsection*{Hypergraph Diagnostics}
We use several methods to extract statistical features from individual hypergraphs and across the set of subjects.

 {\it Hyperedge size:} We define the size, $s(h)$, of a hyperedge $h$, as the number of edges contained in it,
\begin{align}\label{eq:1}
s(h) = \sum_{i,j \in h} \xi^\prime_{i,j},
\end{align} where the sum is performed over the upper triangular elements of $\xi^\prime$, and $\xi^\prime$ is the binarized edge-edge adjacency matrix defined above. Hyperedges with $s(h) = 1$ are singletons, which display no significant correlation between that edge and any other in the network. These singletons are excluded from further analyses. Additionally, we compute the cumulative hyperedge size distribution across all subjects in the study.

{\it Hyperedge node degree:} We define the hyperedge degree of a node to be the number of hyperedges that contain that node. We examine the hyperedge node degree distribution as a spatial distribution over the subjects as a group to understand characteristic hyperedge properties.

\begin{figure}[!ht]
 \centering
 \includegraphics[width = 0.75\textwidth]{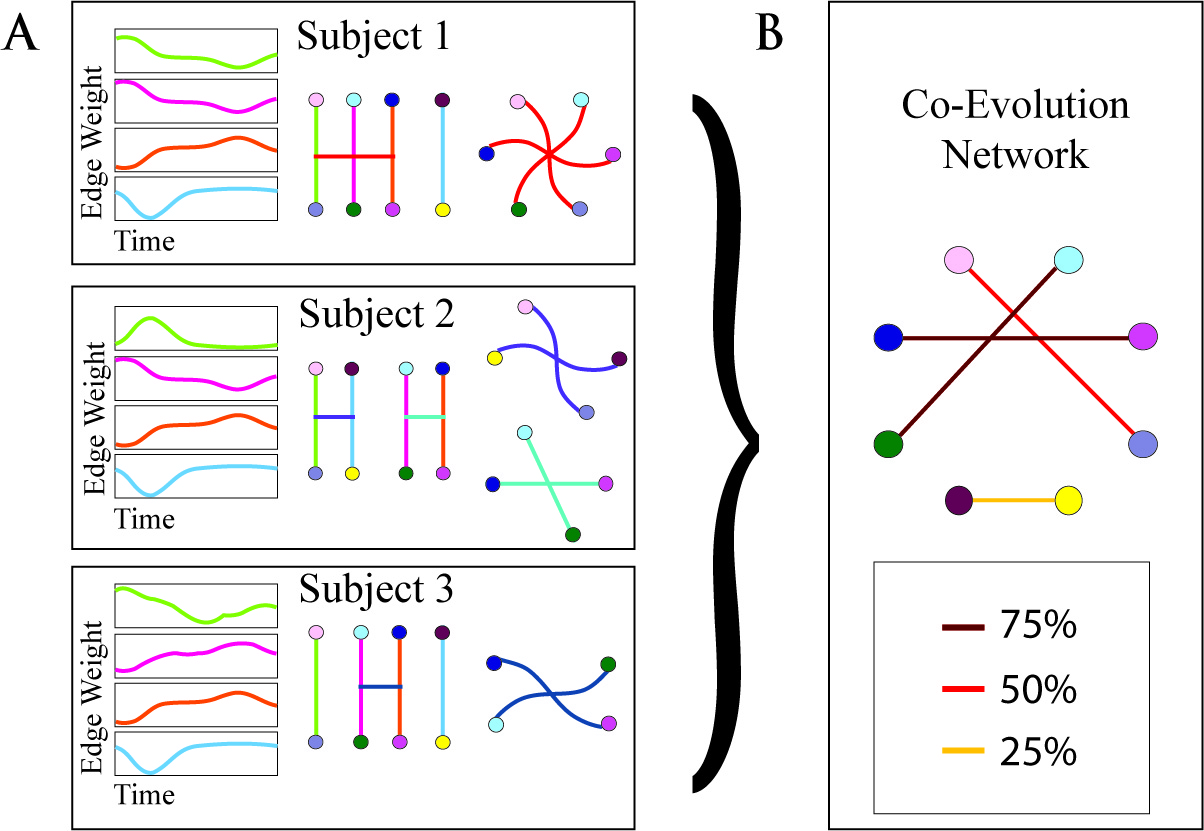}
 \caption{\textbf{Schematic Construction of the Hyperedge Co-Evolution Network:} In (A), we analyze edge time series and group edges exhibiting similar temporal profiles into a hyperedge (as in Figure 1). Here, node colors are used to indicate individual nodes and the edge color indicates distinct edges. We construct hypergraphs for each subject and find the matrix $\textbf{C}$ of probabilities that two nodes are in the same hyperedge over all subjects and hyperedges. In (B), this matrix is used to create a co-evolution network, where the weight for an edge connecting nodes $i$ and $j$ corresponds to the entry $\textbf{C}_{i,j}$.}
 \label{construct}
 \end{figure}

{\it Co-Evolution Network:} We construct a ``co-evolution network" to consolidate hypergraph results into a single graph that illustrates where hyperedges are most likely to be physically located over an ensemble of individuals. Figure \ref{construct} illustrates a schematic of our construction. We begin by defining the matrix, $\textbf{C}$, of probabilities that edges are included in a hyperedge over a set of hypergraphs. Again, nodes correspond to brain regions and connections correspond to inter-region associations, but here the weight of a connection joining nodes $i$ and $j$ is the matrix entry $\textbf{C}_{i,j}$. The resulting static network encompasses the dynamics of hyperedge activity, with connection weight corresponding to the probability that the two nodes are co-evolving over all of the hypergraphs considered. In later sections, we refer to co-evolution connection ``strength," which we define as the magnitude of the probability matrix entry corresponding to that connection.

 \subsection*{Task-Specific Classification}
Previous work identified regions with task-specific activity in rest, attention, and memory tasks \cite{Hermundstad2013a}. Further understanding of the regions that have a correlation structure unique to one task provides insight into network structure differences between tasks. To investigate the task-specific hyperedge structure, we first group hyperedges that exhibit a significantly higher correlation within one task into task-specific sets. If a hyperedge is significantly correlated in two or more tasks, it is excluded from the task-specific hypergraphs. The task-specificity of hyperedges is calculated by comparing the correlation within a single task to the correlation over the same time length with time points chosen randomly from other tasks. This permutation test uses a Bonferroni correction for false positives due to multiple comparisons \cite{hochberg_sharper_1988}. Task-specific hypergraphs are then used to construct task-specific hyperedge size distributions, hyperedge node degree distributions, and co-evolution networks.

To quantitatively probe the differences in spatial organization of dynamic functional co-evolution networks for the four tasks, we investigate two summary metrics that show significant variation across tasks. Choice of these measures is primarily motivated by observed coarse differences in co-evolution network structure.

The first ``length-strength" metric is the Pearson correlation coefficient, $R$, between the strength of a connection in the co-evolution network and Cartesian distance between the two nodes linked by the connection (physical length). The Cartesian distance is computed by taking the $x$, $y$, and $z$ coordinates of each node and calculating the square root of the differences squared. The length-strength metric identifies a geometric property of the network, as well as a coarse estimate of the length of the strongest connections. Furthermore, connection length is related to network efficiency \cite{sporns_small_2004, heuvel_efficiency_2009}, so differences in this measure could indicate varying levels of functional network efficiency corresponding to task states.

The second ``position-strength" metric is the Pearson correlation coefficient, $R$, between the strength of the co-evolution network connection with the average anterior-posterior position of the two nodes. A measure of anterior-posterior position for each connection was found by taking the average $y$ position of the two nodes in the connection. Identifying the location of strong co-evolution network connections along the anterior-posterior $y$ axis provides a measure of where hyperedges are physically present in task states. Both the structural core \cite{hagmann2008mapping} and a dynamic functional core area, comprised of sensorimotor and visual processing areas \cite{Bassett2013a}, are located in the posterior, so nodes in these regions have negative $y$ values. A larger negative position-strength value corresponds to a higher probability that hyperedges are active in these core areas.

The length-strength and position-strength metrics are evaluated for significance by comparing the correlation between length or position and connection strength to the same correlation performed on randomly chosen co-evolution connections. Again, the Bonferroni correction is performed to eliminate false positives due to multiple comparisons. 

In Results, we discuss how these metrics reveal quantitative differences between task-specific networks. A more detailed analysis of the overlap between hyperedge co-evolution networks and relevant cognitive processing regions is also presented. In this analysis, we describe how delineated areas of higher hyperedge activity consistently correspond to recognized centers of task-specific activity.

\subsection*{Null Models}

In this analysis, we compare our results with two statistical null models based on measures for dynamic networks \cite{Bassett2012b}. Hyperedges are formed from correlated edge time series; consequentially the null overall model randomly shuffles each edge time series over all experiments. This null model is designed to ensure that the hyperedges identified in our analysis can be attributed to the dynamics of the system, rather than some overall statistical property of the data set.

The other null test we perform, which we will refer to as the null within-task model, reorders each edge time series within each task, keeping tasks distinct. This is constructed in order to determine whether there are specific differences in the data between tasks.

\section*{Results}
 We compile the results from the hypergraph analysis for each of the subjects and combine these results to obtain a size distribution, anatomical node degree distribution, and co-evolution network for the group. We then divide the data into task-specific hypergraphs and perform the previously mentioned analyses on the task-specific hypergraphs.

\subsection*{Hypergraph Analysis and Statistics}

\begin{figure}[!ht]
 \centering
 \includegraphics[width = 0.5\textwidth]{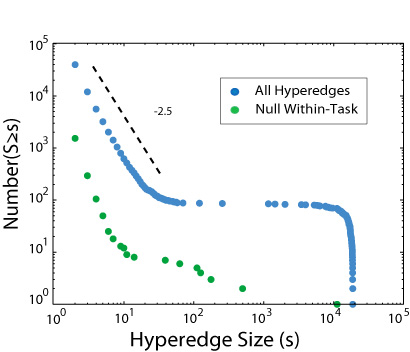}
 \caption{\textbf{Hyperedge Size Distribution:} In the cumulative frequency distribution of hyperedge sizes, the small hyperedges appear to roughly follow a power law with an exponent of approximately $-2.5$, while the large group is concentrated near the maximum size. In the null overall model, there are no non-singleton hyperedges. Results for the null within-task model, where the data is shuffled within each task, are in green.}
 \label{Figure 2}
 \end{figure}

We construct a hypergraph for each individual and examine the cumulative distribution of hyperedge sizes ($s(h)$ from Equation \ref{eq:1}), shown in Figure \ref{Figure 2}. There is a distinct break in the slope between two branches of the distribution occurring at a size of approximately 100 edges, which we use to distinguish between ``large" and ``small" hyperedges. The total number of small hyperedges appears to roughly follow a power law with an exponent of approximately $-2.5$. The number of large hyperedges peaks around the maximum size, with relatively few in the middle range from 100 to 1000 edges.  In Figure \ref{Figure 2}, the sharp drop off in the distribution at large hyperedge sizes reflects the system size limitation on hyperedge cardinality.
 
 There is a distinct partition in all individual frequency versus sizes distributions; one or two ``large" hyperedges ($s(h) > 100$), and many ``small" hyperedges that peak at the smallest size ($s(h) < 100$). For purposes of illustration, the individual hypergraph of a subject with a relatively small maximum hyperedge size is shown in Figure \ref{HI}. The corresponding hypergraph of a subject with a maximum hyperedge near the system size is strongly dominated by the largest hyperedge, which obscures the smaller hyperedges.
 
  The null overall model shuffles the data over all tasks. There are no hyperedges greater than size one, so the results from this null model are not depicted in Figure \ref{Figure 2}. These singletons signify no significant correlation with other edges. As a result, we performed no further analysis on this null model. The fact that no significant hyperedges were found in the null overall model validates the statistical significance of our results.
 
  The null within-task model shuffles the data but ensures that task data stays within the same task. The size distribution of hyperedges from the null within-task model is shown in Figure \ref{Figure 2}. The shape of the two distributions is similar, although the null within-task model has fewer hyperedges in the large regime and there are more singletons than in the original data. This indicates there is co-evolution structure across tasks because this structure corresponds to changes in edge states between two or more tasks. For example, if groups of edges have an overall high correlation in one task and a significantly lower correlation in another, it would induce a hyperedge across the tasks regardless of how the within-task time series are shuffled.

\begin{figure}[!ht]
 \begin{center}
 \includegraphics[width = 0.5\textwidth]{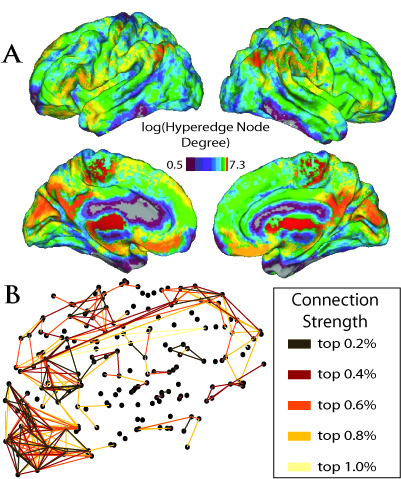}
 \end{center}
  \caption{
  {\bf Hyperedge Node Degree and Co-Evolution Network:} In (A), we show hyperedge node degree on a natural log scale. The cumulative number of hyperedges at each vertex over all individuals is plotted on the brain, where higher values at a node correspond to more hyperedges that include the node. (B) depicts a sagittal view of the co-evolution network. The edge strength represents the probability that the edge will be in a hyperedge over all individuals. Edge color corresponds to threshold percentage value, where only the top 1\% of co-evolution probabilities are shown. Within this 1\%, brown connections correspond to the highest 0.2\% of probabilities, red connections correspond to 0.2\% to 0.4\%, orange connections correspond to 0.4\% to 0.6\%, gold connections correspond to 0.6\% to 0.8\%, and yellow connections correspond to 0.8\% to 1\%.}
  \label{Figure 4}
  \end{figure}

 Examining the cumulative hyperedge size distribution provides information about the network topology but does not supply descriptive spatial information. Next, we quantify which anatomical locations in the brain participate in hyperedges, identifying differential roles in task-induced co-evolution. Figure \ref{Figure 4}A depicts the hyperedge node degree on a natural log scale. The densest regions are located in posterior portions of the cortex, primarily in visual areas, while a second set of dense regions is located in the prefrontal cortex.
 
We construct a co-evolution network, as illustrated schematically in Figure \ref{construct}, where connection weight corresponds to the probability that two nodes participate in the same hyperedge. In Figure \ref{Figure 4}B we present this co-evolution network over all individuals and all tasks. The graph includes sparse long-range connections between regions that are densely connected. Within the strongest 1\% of connections, the high degree of bilateral symmetry indicates that corresponding nodes in the left and right hemispheres have a high likelihood of being placed together in a hyperedge. Dense areas of the graph include primary visual areas, portions of prefrontal cortex, and primary motor cortex.

\subsection*{Task-Specific Hyperedges}
\begin{figure}[!ht]
 \begin{center}
 \includegraphics[width = 0.75\textwidth]{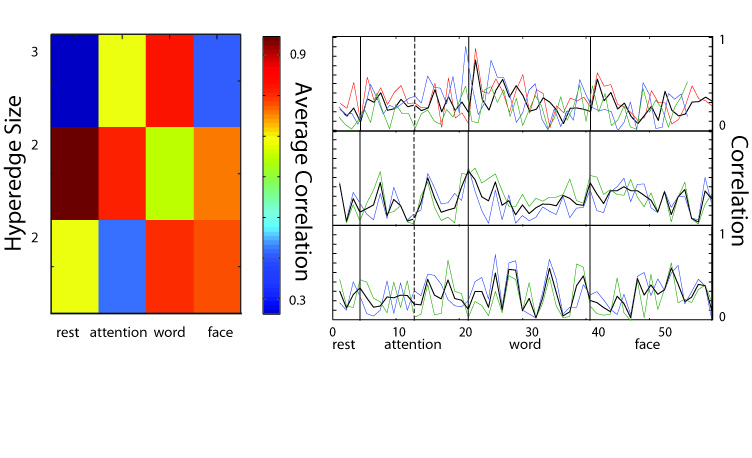}
 \end{center}
 \caption{
 {\bf Left:} Average hyperedge correlation in each task for three hyperedges (where hyperedges with small sizes are chosen for illustrative purposes). {\bf Right:} Correlation (absolute value) time series for the same three hyperedges. The colored lines represent each edge, while the black line is the average edge time series. Each time point represents the static network over 60 seconds, and the attention task is broken into two sections because two separate iterations of the same task were combined in this analysis. These results display the task-specificity of hyperedges, where significant correlations in the hyperedge are restricted to one task. For example, the first hyperedge is word-specific because there is a much stronger average correlation in the word task than in any other task.}
 \label{tasksp}
 \end{figure}

The hypergraph algorithm groups together edges with significantly similar temporal behavior. However, this basic classification does not distinguish whether the correlation is present throughout the edge time series, or whether highly correlated sections of the time series drive the selection. We compute the average within-task edge correlation for each hyperedge and find that in some cases, strong edge correlation spans the tasks, while in other hyperedges, a strong correlation between edges within one task drives the hyperedge. An example of this task-specific correlation structure can be seen in Figure \ref{tasksp}. In the average within-task correlation on the left, there is a stronger average correlation in the word memory task than in any other task. Furthermore, the edge time series in the first hyperedge indicates it is driven mainly by a correlation within the word memory task.

To investigate this further, we construct task-specific co-evolution networks, composed of hyperedges with significantly stronger average correlation in one task than the others (see Methods). To identify these task-specific hyperedges for each task, we perform a permutation test on the edge weight time series, as described in Methods, and compare the total correlation within the task to the expected values. If a hyperedge displays significant edge correlation (determined by the Bonferroni correction on the $p$-values from the permutation test) in only one task, we label it as a task-specific hyperedge. Hyperedges with two or more tasks exhibiting significant correlation are not included in the task-specific hypergraphs.

\begin{figure}[!ht]
\begin{center}
\includegraphics[width = 0.75\textwidth]{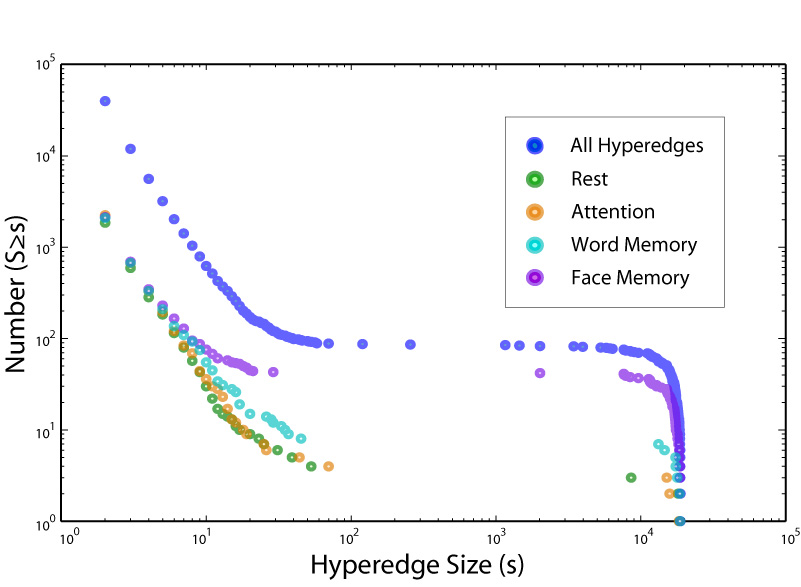}
\end{center}
\caption{
{\bf Task Specific Hyperedge Size Distributions:} Cumulative frequency distribution as a function of hyperedge size for all task-specific groups. The results are compared to the overall distribution of hyperedges (dark blue), previously illustrated in Figure \ref{Figure 2}. There are fewer large hyperedges attributed to attention and rest tasks, while the memory tasks have a greater number of large task-specific hyperedges.}
 \label{Figure 5}
 \end{figure}

Figure \ref{Figure 5} illustrates the size distributions of all the task-specific results alongside the overall hyperedge size distribution. The sizes and spatial distributions of single task-driven hyperedges vary across tasks and incorporate significant information about functional network organization with respect to changing cognitive states. Attention has the greatest number of task-specific hyperedges, followed by face memory, word memory, and rest. In the small regime, the tasks follow a similar distribution. There are fewer large attention and rest hyperedges, while the face memory task closely mimics the overall distribution. The distinction in the distributions indicates that the tasks can be characterized by differing complexities of edge co-variations.

\begin{figure*}[!ht]
\begin{center}
\includegraphics[width=\textwidth]{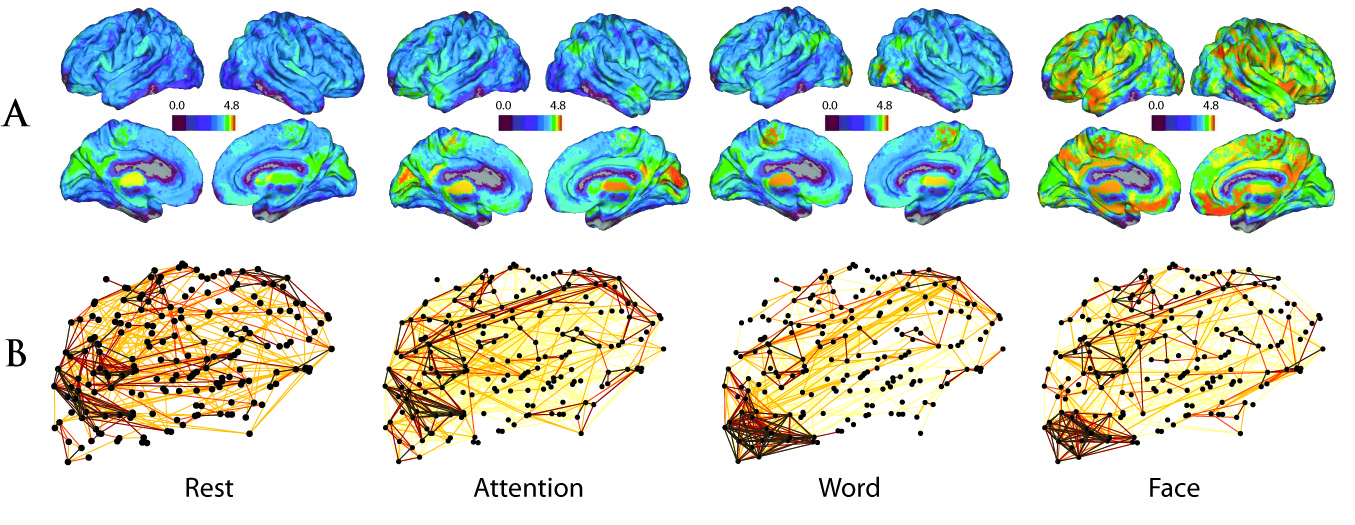}
\end{center}
\caption{
{\bf Task-specific Co-Evolution Networks and Hyperedge Node Degrees:} \\    
(A): Distribution of task-specific hyperedge node degree on the brain. Here, the log of the total number of hyperedges containing each node is represented on the brain. The color scale represents the log of hyperedge node degree as in \ref{Figure 4}A, although here the range of values is from 0 to 4.8. (B): Co-evolution networks for each task. Edge strength corresponds to the probability that a hyperedge will contain the edge over all individual hypergraphs. Color represents a threshold in percentage value, with the scale given in Figure \ref{Figure 4}B, and the top 1\% of co-evolution probabilities are shown. Once again, the top 2 \% of probabilities are brown, red indicates the top 0.2\% to 0.4\% of connections, orange indicates the top 0.4\% to 0.6\% of probabilities, gold indicates the top 0.6\% to 0.8\% of probabilities, and yellow indicates the top 0.8\% to 1\% of probabilities.}
\label{Figure 6}
\end{figure*}

\begin{figure}[!ht]
\begin{center}
\includegraphics[width = 0.75\textwidth]{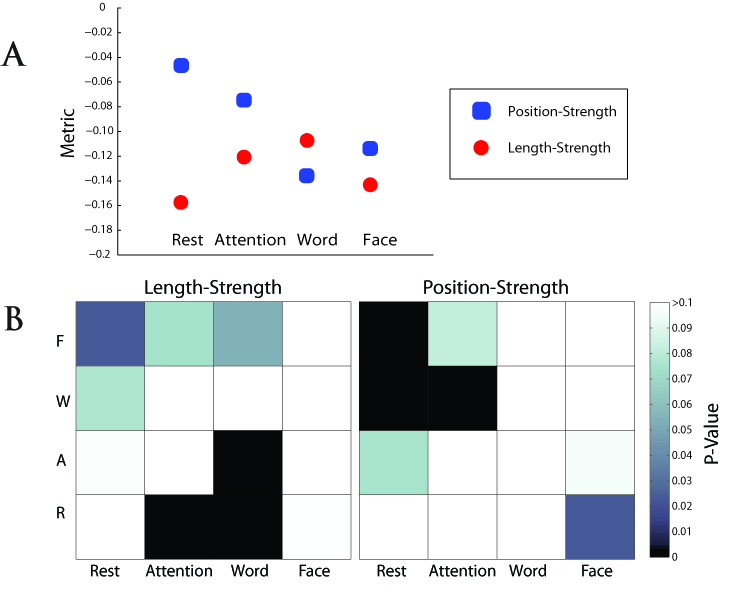}
\end{center}
\caption{
{\bf Task-specific network statistics:} Values for the position-strength metric (blue) and the length-strength metric (red) for the four tasks are depicted in (A). (B) shows $p$-values for the pairwise statistical permutation test between tasks, where black denotes a significant value after a Bonferroni correction for multiple comparisons. Values are obtained for length-strength and position-strength metric. For example, on the $y$ position plot in (B), attention-word is significant. Referring back to (A), we see that this implies the difference in the $y$ position-strength correlation between the attention and word tasks is statistically significant.}
\label{Figure 7}
\end{figure}

The spatial distributions of hyperedge node degree in each task, along with task-specific co-evolution networks, are shown in Figure \ref{Figure 6}. The rest hypergraph has the least activity in posterior regions of the cortex, both in the hyperedge node degree plot and co-evolution network. In the attention network, long connections connecting the front and back of the brain distinguish it from the rest network. Furthermore, the concentration in the occipital lobe is larger in the memory co-evolution networks than in the rest or attention networks. We characterize these observed differences with two statistics, which are described in more detail in Methods. The length-strength metric is a correlation between connection length and strength in the co-evolution network. The position-strength metric is a correlation between connection position (anterior-posterior) and strength. The results of this analysis over the full unthresholded co-evolution network are in Figure \ref{Figure 7}. All correlation values are negative, indicating that, in all tasks, stronger connections in the co-evolution network are located in posterior portions of cortex and are physically shorter.

We compare these values across tasks by performing pairwise permutation tests to determine which networks have statistically different properties. Figure \ref{Figure 7} depicts the $p$-values from these tests, where the horizontal axis represents the statistic being tested and the vertical axis corresponds to the task being tested against. The black squares in this figure represent significant values, which are summarized in the following list:

\begin{enumerate}
\item The rest task has a significantly less strong position-strength correlation than the word and face memory tasks. This confirms the observation that the rest co-evolution network is less likely than the memory networks to have strong connections in posterior regions of the cortex.

\item The attention task is less strongly correlated than the word memory task, as measured by the position-strength metric and the rest task in terms of the length-strength metric. Thus, the attention co-evolution network is less likely than word memory to have strong connections in the posterior, and less likely than the rest network to have strong connections that are short.

\item The word memory task has a weaker length-strength correlation than the rest and attention tasks. Thus, strong connections in the word memory co-evolution network are less likely be short than they are in attention and rest networks.
\end{enumerate}

These results delineate significant differences in co-evolution network structure between the tasks, confirming that the hypergraph analysis is a useful method for distinguishing between task states. Additional features of the task-specific co-evolution networks are described in more detail below.

\subsubsection*{Rest}

Rest-specific hyperedges are primarily represented in the ``small" range of the size distribution in Figure \ref{Figure 5}. Additionally, the rest task has the lowest number of task-specific hyperedges, but it is difficult to distinguish this in Figure \ref{Figure 5} due to the proximity on the logarithmic scale. The result that fewer groups of nodes share similar co-evolution properties could be a result of the specificity of correlated resting state regions, or a simplicity intrinsic to resting state function that does not necessitate more concerted efforts involving numerous brain regions \cite{cole_intrinsic_2014}.

There are fewer rest-specific hyperedges than in any other task, so the hyperedge node degree plot in Figure \ref{Figure 6}A has the lowest overall magnitude across task states. The areas with the highest degree of hyperedge activity are in the posterior portions of the brain, a configuration that is consistent across tasks. This suggests there is an underlying pattern of hyperedge generation centered in the occipital lobe.

The rest co-evolution network is highly clustered in the most probable 0.2\% of co-evolution pairs, but lower thresholds show very little structure, as visualized in Figure \ref{Figure 6}B. These high probability clusters are centered in previously mentioned regions, but the top 1\% of connections is far more randomized in rest than any other task-specific co-evolution network we have observed. There is relatively little lateral symmetry and few visible ``core" areas with high hyperedge node degree.

We find the negative length-strength correlation is significantly stronger for the rest task than the word memory task. This indicates that the strongest connections in the co-evolution network for the rest task are short, reflecting the initial observations in Figure \ref{Figure 6}B. The rest co-evolution network has the smallest negative correlation between connection position and strength, which the permutation test (Figure \ref {Figure 7}B) confirmed to be significantly smaller than the word or face memory tasks. This relative lack of long, strong connections may correspond to a diminished need for efficient processing in a task-free environment.

Resting-state brain activity contains correlated patterns that comprise a default mode network, a system that is engaged during internal cognition \cite{long_default_2008,albert_resting_2009}. Certain brain regions active at rest are consistently deactivated during goal-oriented tasks, indicating that they comprise a functional mode that is rest-specific \cite{Raichle2001}.

Our analysis returns a spatial co-evolution distribution with structures of higher probability in brain regions traditionally associated with the resting state. Dense areas of the network with high probabilities of being in the same hyperedge include the inferior parietal lobule, superior frontal gyrus, precuneus, and posterior cingulate cortex. These regions have been identified as integral components of the default mode network; the posteromedial cortex includes the precuneus and posterior cingulate cortex and performs a particularly pivotal role in awareness and memory retrieval \cite{dastjerdi_differential_2011,fransson_precuneus/posterior_2008,cauda_functional_2010}. Moreover, as in all of the tasks, the rest task has the greatest concentration of hyperedges in the posterior of the brain. This may overlap with previously located structural and functional core areas and is an avenue for further investigation.

\subsubsection*{Attention}

The attention-specific network primarily consists of small hyperedges. Overall, there are more hyperedges associated with attention than any other task, but this is difficult to visualize in Figure \ref{Figure 5}. This composition of many small hyperedges may account for the increased disorganization in the co-evolution structure at lower probability thresholds.

In Figure \ref{Figure 6}A, the hyperedge node degree plot for the attention task appears qualitatively similar to the rest task, with a few areas of increased degree in the occipital lobe. The hyperedge node degree values are larger here, corresponding to the greater number of hyperedges that are present in the attention task.

In Figure \ref{Figure 6}B, the co-evolution structure specific to the attention task is depicted. The structure of the attention network has a higher degree of bilateral symmetry than the rest network, but has significantly fewer connections in the occipital lobe than either memory task. There are multiple prefrontal cortical regions that are likely to cohesively evolve with several other nodes. Numerous links between rostral and caudal brain regions are another unique feature of the attention network.

The negative length-strength correlation in the attention co-evolution network is significantly less strong than in the rest task, as we can verify in Figure \ref{Figure 6}B by observing that the attention co-evolution network has strong connections that reach across the brain. Additionally, the attention task has a weaker position-strength correlation than the word memory task. It is likely that the strong connections in the prefrontal cortex are driving this feature, which could indicate that the attention task state has less reliance on core visual regions than the word memory task.

Two attention systems exist in the human brain: a ``top-down" network controls goal-directed attention, while a ``bottom-up" group of brain regions detect and orient attention to relevant sensory stimuli that are generally novel or unexpected \cite{corbetta_control_2002, fox_spontaneous_2006}. Our task probes the former, as subjects are asked to focus on repetitive stimuli in a controlled environment. This requires an ``executive control network," a bilateral dorsal system that governs guided attention and working memory \cite{seeley_dissociable_2007}.

Regions of high clustering in the most probable threshold include the lateral parietal occipital lobe, the superior frontal cortex, and dorsal parietal cortex. The parietal and frontal areas are involved in attention control and localization, specifically in visual attention tasks \cite{corbetta_control_2002,nobre_functional_1997}. Activation of the superior frontal cortex occurs in attention tasks, but a significant increase in activity occurs during tasks where peripheral attention is required \cite{corbetta_pet_1993, hopfinger_neural_2000}. The dorsal parietal cortex also performs a central role in the executive control network; patients with lesions in the dorsal parietal cortex have shown significant impairment in goal-directed attention tasks \cite{shomstein_goal-directed_2005}.

\subsubsection*{Memory for Words}
The word memory-specific hyperedge size distribution includes numerous large hyperedges, but is not as close to the overall distribution as the face memory distribution. Together, the memory tasks contain many more large hyperedges than the rest and attention tasks. This indicates some aspect of the memory tasks requires dynamically coherent evolution over much of the brain.

The word memory hyperedge node degree distribution has high node degrees in similar areas to the other task-specific hypergraphs. There is a marked increase in node degree of regions in the parietal lobe from rest and a decrease in degree of regions in the occipital lobe from attention (seen in Figure \ref{Figure 6}A).

Visual orthographic and face processing have a common reliance on central vision \cite{levy_center-periphery_2001} and share neural circuitry \cite{nestor_neural_2013}. The resemblance of the co-evolution networks for the two tasks, especially when compared with the very different graph structure of the attention and rest networks, indicates a similarity in the hypergraph representation of the memory tasks. This in turn signifies a correspondence in brain dynamics specific to memory. The task-specific analysis requires hyperedges to show a significant correlation in only one task, so there is no overlap in these co-evolution networks.

The negative length-strength correlation is the weakest for the word memory network. We find that these two variables are significantly less correlated than in the rest or face memory tasks.  This indicated that the many connections from the occipital to frontal lobes are a distinguishing characteristic of the word co-evolution network.

Existence of a dedicated visual word processing network has been a topic of frequent discussion in neuroscience.  The visual word form area (vWFA), located in the occipito-temporal cortex, is consistently activated by orthographic stimuli \cite{turkeltaub_meta-analysis_2002} and is invariant to changes in case, size, font, or type of visual stimulation \cite{polk_functional_2002,rauschecker_visual_2011}. The vWFA has also been shown as functionally linked to the dorsal attention network in resting state fMRI data, indicating that it fulfills a complex cognitive role \cite{vogel_putative_2011}.

In the word memory co-evolution network in Figure \ref{Figure 6}B, low strength co-evolution pairs are more broadly distributed throughout the brain, while the strength and number of bilateral links is diminished. This is consistent with the understood activation structure of working memory tasks \cite{klingberg_bilateral_1997}. The vWFA is highly connected in the co-evolution network, but there is minimal strong structure in dorsal attention areas, which we would expect to see in a functional connectivity analysis \cite{vogel_putative_2011}. This can be explained by our methodology of selecting task-specific hyperedges. If edges in the dorsal attention network have similar co-evolution properties within the word memory and attention tasks, they will not be identified as task-specific edges.

\subsubsection*{Memory for Faces}

There are more large hyperedges significantly correlated in the face memory task than any other task-specific group. The distribution closely resembles the overall distribution in the large regime, indicating that a significant portion of these large hyperedges are driven by correlations in the face memory task.

The face memory-specific hyperedge node degree values are consistently the largest across the brain. In the word memory and attention degree distributions, there are areas of higher hyperedge node degree in the parietal lobe and occipital lobe, respectively, but the face memory degree distribution is more evenly dispersed over the brain. This is primarily due to the many large hyperedges specific to the face memory recognition task.

The structure of the co-evolution network, in Figure \ref{Figure 6}B, is most dense in the occipital lobe, which corresponds to the visual nature of the task. There are several strong connections from the occipital lobe to other brain regions, specifically in the frontal lobe. While the structure looks similar to the word memory task, the strong cluster of connections in the occipital lobe have fewer strong connections and less nodes reached overall, but more strong connections to a few particular nodes in the face memory task, and there are fewer strong connections in the frontal lobe but more strong connections among regions in the dorsal attention network. 

In addition to the properties discussed in previous sections, the face memory co-evolution network has a strong negative position-strength correlation. The structure at the highest probability levels of the co-evolution network is primarily contained in the prefrontal cortex and areas known to be structural and functional core regions \cite{hagmann2008mapping, Bassett2013a}.

Face recognition in humans requires a complex network distributed throughout the visual cortex that includes extended connections branching to other cortical regions \cite{haxby_human_2002}. The majority of visual processing occurs in the occipital lobe, located in the posterior of the brain. Functional MRI studies have identified multiple regions in the occipital cortex that respond more strongly to faces than other visual stimuli, indicating that the cognitive processes involving facial recognition are highly specialized \cite{kanwisher_fusiform_1997,gauthier_fusiform_2000}. A region in the fusiform gyrus, the fusiform face area (FFA), is selectively active in whole human facial perception \cite{mccarthy_face-specific_1997,saygin_anatomical_2012}.

The face perception system is composed of multiple bilateral regions; lateral symmetry in the co-evolution network is consistent with this structure \cite{haxby_human_2002}. An aspect of the network that breaks this symmetry is the right fusiform gyrus, which is strongly connected to other areas in the occipital lobe by high probability co-evolution pairs. In agreement with evidence that many aspects of facial perception are right-hemisphere dominant, the right FFA has the most salient response to faces; damage to the region severely impairs face recognition \cite{saygin_anatomical_2012}.

\section*{Discussion}
Progress in understanding functional brain network topology provides significant insight into broad neuroscience questions regarding the brain's organization and ability to effectively transition between cognitive states. Quantifying complex network dynamics in the brain will further understanding in these areas and has promising applications to behavioral adaptation and learning \cite{Bassett2011b, mantzaris_dynamic_2013,Bassett2013a}. We apply hypergraph analysis, a tool from dynamic network science, to functional brain imaging data in order to determine co-evolution properties of the brain as subjects perform a series of tasks. A previous application of this method to neuroscience uses hypergraphs to analyze how functional network structure changes over a long term learning task \cite{Bassett2013a}. The learning experiment considers hypergraphs constructed over 6 weeks of training while subjects acquire a new motor skill, while our analysis compares hypergraphs over three different tasks performed within an interval of hours. Our analysis shows that hypergraphs are a useful tool for investigating shorter time scales and differentiating between task-specific networks.

Instead of analyzing the time-dependent behavior of groups of nodes, the hypergraph investigation considers the edge weight time series, where edges with statistically significant similarities in their temporal profiles are grouped into hyperedges. This approach is advantageous because it considers all edges, regardless of correlation strength, unlike previous methods which focus exclusively on strong correlations \cite{Hermundstad2013a}. The use of a data-driven analysis allows us to investigate the dynamic changes in brain function over a series of tasks without prior assumptions of the structure of the connectivity network. This is a significant advantage over methods that characterize task states based on their differences with respect to the resting network \cite{Fox2005,Hampson2010}. A comparison between the hypergraph analysis and these methods in a future analysis could reveal how the concentration of hyperedges varies in known task-positive or task-negative areas and determine whether this variation has an effect on task performance.

\subsection*{Hypergraph Statistics and Structural Metrics}

We demonstrate the existence of hypergraph structure in functional brain dynamics and statistically characterize the hyperedge distributions in comparison to appropriate null models. Shuffling the time series over all time produces no significant hyperedges, while shuffling within each task results in a size distribution that resembles the overall size statistics with far fewer hyperedges.  The distinct differences between the two null models and our results based on the original time series establish the significance of our findings. Furthermore, the existence of hyperedges after the within-task shuffling indicates the presence of activity in some edges that is differentiated between tasks. Since there are fewer large hyperedges after the within-task shuffling, we can also confirm that there are hyperedges caused by edge dynamics within tasks. This work primarily concentrates on hyperedges correlated within a particular task, but future analyses to understand the properties of hyperedges that are grouped due to other general properties would supplement our results.

The hyperedge size distribution is comprised of ``small" and ``large" hyperedges, where the size distribution of the small hyperedges follows a power law and the large hyperedges peak at the system size. We explore the overall spatial hyperedge distribution by constructing a hyperedge node degree plot, and find that the majority of the most densely connected nodes lie in the posterior portions of the brain. To better observe spatial hyperedge properties, we develop a co-evolution network, where connection weights correspond to the probability that a hyperedge will include the connection. The top 1\% of connections in the network with the highest probability of inclusion in a hyperedge are most concentrated in the occipital lobe and prefrontal cortex. These are expected areas of hyperedge concentration, consistent with the visual nature of the tasks, as well as the coordination of quick decision making and the selection of specific motor responses.

\subsection{Task Specificity and Anatomical Placement}
We find there are hyperedges that are more correlated in one task and hyperedges that have a distinct profile across the tasks. Our results suggest that edges with a high probability of inclusion in task-specific hyperedges are often found in previously identified brain areas associated with the corresponding tasks, confirming that the approach captures relevant information about task networks. In some networks, brain regions expected to show strong co-variation in a certain task are not present in the co-evolution networks. Repeating the analysis and grouping edges that are significantly correlated in two tasks might lend insight into whether brain systems relevant to a certain task contain hyperedges that are correlated in another task and thus are rejected from our task-specific analysis.

In all tasks, stronger connections in the co-evolution network tend to be located in posterior portions of cortex and to be physically shorter. The higher probability of posterior edges to be included in hyperedges is consistent with the identification of a core set of highly structurally connected regions centered in the posterior of the brain, thought to play an important role in integrating large-scale functional connectivity \cite{hagmann2008mapping}. The tendency of strong connections to be physically shorter suggests high efficiency in task-specific co-evolution networks. This may reflect efficient wiring properties associated with minimal wiring for rapid processing and low energy expenditures found in structural brain networks and shared by some other biological and technological networked systems \cite{Bassett2010}.

The co-evolution networks for memory tasks show a significantly higher hyperedge probability in visual areas than the attention and rest tasks, and the differences in structure indicate that the hypergraph representation of memory tasks is significantly different from rest or attention. The marked differences in hyperedge statistics in our task-specific analysis suggest hypergraphs as a measure of functional network changes due to task states. With measures derived from the hyperedge analysis, we can begin to quantitatively probe the mechanisms of functional switching between tasks and gain insight into how distinct features of the network evolve in synchronized patterns.

\subsection*{Final Remarks}
Efforts to make quantitative comparisons between the hypergraph analysis and other dynamic graph theoretical methods in the context of the human brain are ongoing. We are investigating whether dynamic community detection, a node-based analysis, can provide complementary information to the edge-based hypergraph analysis. Throughout this work, we also observe a significant amount of individual variability in the hypergraph properties of interest. This variability may be related to differences in cognitive ability, which is an avenue we are investigating further.

In this paper, we use hypergraph analysis to identify significant co-evolution between brain regions in task-based functional activity and develop new tools to summarize the spatial patterns of these co-evolution dynamics over the group of subjects. By isolating task-specific hyperedges, we quantify significant differences between the spatial organization of co-evolution dynamics within different tasks. This hypergraph analysis adds a crucial perspective to previous treatments of task-based brain function, describing temporal similarities between spatially segregated neural circuits by specifically examining the organization of connections that co-evolve in time. It provides a promising approach for understanding fundamental properties of task-based functional brain dynamics, and how individual variation in these properties may correspond to differences in behavior and task performance.

\section*{Acknowledgments}
We would like to thank John Bushnell for technical support and Ben Turner for invaluable help with visualization. This work was supported by the David and Lucile Packard Foundation and the Institute for Collaborative Biotechnologies through grant W911NF-09-0001 from the U.S. Army Research Office. K.J.S. was supported by the National Science Foundation Graduate Research Fellowship Program under Grant No. DGE-1144085. E.N.D. and K.J.S. were additionally supported by the Worster Fellowship. D.S.B. was supported by the Alfred P. Sloan Foundation, the Institute for Translational Medicine and Therapeutics at Penn, and the Army Research Laboratory through contract number W911NF-10-2-0022. The  content of the information does not necessarily reflect the position or the policy of the Government, and no official endorsement should be inferred.

\bibliography{bibfile}

\end{document}